\documentclass[longauth]{aa}
\usepackage{txfonts}
\usepackage{graphicx}
\usepackage{natbib}
\usepackage{color}
\bibpunct{(}{)}{;}{a}{}{,}

\def\DS {D_\mathrm{S}}
\def\DL {D_\mathrm{L}}

\def\ML {M_\mathrm{L}}
\def\to {t_{\circ}}
\def\uo {u_{\circ}}
\def\tE {t_\mathrm{E}}
\def\ThE {\theta_\mathrm{E}}
\def\piE {\pi_\mathrm{E}}
\def\RS {R_\mathrm{*}}
\def\ThS {\theta_\mathrm{*}}
\def\rhoS {\rho_\mathrm{*}}
\def\msun {M_{\odot}}
\def\lsun {L_{\odot}}
\def\rsun {R_{\odot}}

\def\teff {T_\mathrm{eff}}

\def\Vs {V_\mathrm{s}}
\def\Is {I_\mathrm{s}}
\def\VI{(V-I)}
\def\VIo {(V-I)_{\circ}}
\def\VKo {(V-K)_{\circ}}
\def\JKo {(J-K)_{\circ}}
\def\JHo {(J-H)_{\circ}}
\def\HKo {(H-K)_{\circ}}

\def\Vo {V_{\circ}}
\def\Io {I_{\circ}}
\def\Ko {K_{\circ}}
\def\Ks {K_\mathrm{s}}

\def\Ai {A_\mathrm{I}}
\def\Av {A_\mathrm{V}}
\def\Aj {A_\mathrm{J}}
\def\Ah {A_\mathrm{H}}

\def\Aks {A_\mathrm{Ks}}

\def\logg {\log g}
\def\logl {\log L/\lsun}

\def\Mbol {M_\mathrm{bol}}

\def\Mi {M_\mathrm{I}}
\def\Mj {M_\mathrm{J}}

\def\chisq {\chi^2}
\def\chidof {\chisq/\mathrm{d.o.f.}}
\def\muas {\mu \mathrm{as}}

\def\vperp {v_{\perp}}
\newcommand\Sec[1]  {Sec.~\ref{#1}}
\newcommand\Equ[1]  {Eq.~(\ref{#1})}
\newcommand\Fig[1]  {Fig.~\ref{#1}}
\newcommand\Tab[1]  {Table~\ref{#1}}
\begin{document}
  \title{OGLE~2008--BLG--290: An accurate measurement of the limb darkening of a Galactic Bulge K Giant spatially resolved by microlensing}
  \titlerunning{OGLE~2008--BLG--290, a microlensed Galactic Bulge K Giant}
  \authorrunning{P.~Fouqu\'e, D.~Heyrovsk\'y, S.~Dong \emph{et al.}}
  \author{P.~Fouqu\'e\inst{1,44} \and D.~Heyrovsk\'y\inst{48} \and S.~Dong\inst{2,7}
\and A.~Gould\inst{2,7} \and A.~Udalski\inst{3,27} \and M.D.~Albrow\inst{1,8}
\and V.~Batista\inst{1,11} \and J.-P.~Beaulieu\inst{1,11} \and D.P.~Bennett\inst{1,4,15}
\and I.A.~Bond\inst{4,30} \and D.M.~Bramich\inst{1,5,10} \and S.~Calchi~Novati\inst{50}
% ...
}
  \institute{
    Probing Lensing Anomalies Network, http://planet.iap.fr 
    \and Microlensing Follow Up Network, http://www.astronomy.ohio-state.edu/~microfun 
    \and The Optical Gravitational Lensing Experiment, http://ogle.astrouw.edu.pl 
    \and Microlensing Observations in Astrophysics, http://www.phys.canterbury.ac.nz/moa 
    \and Robotic Telescope Network, http://robonet.lcogt.net 
    \and Microlensing Network for the Detection of Small Terrestrial Exoplanets, http://www.mindstep-science.org 
    \and Department of Astronomy, Ohio State University, 140 West 18th Avenue, Columbus, OH 43210, USA 
    \and University of Canterbury, Department of Physics \& Astronomy, Private Bag 4800, Christchurch 8020, New Zealand 
    \and European Southern Observatory (ESO), Casilla 19001, Vitacura 19, Santiago, Chile 
    \and European Southern Observatory (ESO), Karl-Schwarzschild-Stra\ss{}e 2, 85748 Garching bei Mu\"nchen, Germany 
    \and Institut d'Astrophysique de Paris, CNRS, Universit\'{e} Pierre \& Marie Curie, 98bis Bd Arago, 75014 Paris, France 
    \and Astronomisches Rechen-Institut (ARI), Zentrum f\"{u}r Astronomie der Universit\"{a}t Heidelberg (ZAH), M\"{o}nchhofstrasse 12­-14, 69120 Heidelberg, Germany 
    \and Scottish Universities Physics Alliance, School of Physics \& Astronomy, University of St Andrews, North Haugh, St Andrews, KY16~9SS, United Kingdom 
    \and University of Notre Dame, Department of Physics, 225 Nieuwland Science Hall, Notre Dame, IN 46556, USA 
    \and University of Texas, McDonald Observatory, 16120 St Hwy Spur 78, Fort Davis TX 79734, USA 
    \and Institute of Geophysics and Planetary Physics (IGPP), L-413, Lawrence Livermore National Laboratory, P.O. Box 808, Livermore, CA 94551, USA 
    \and Physics Department, Faculty of Arts and Sciences, University of Rijeka, Omladinska 14, 51000 Rijeka, Croatia 
    \and Technical University of Vienna, Dept. of Computing, Wiedner Hauptstrasse 10, Vienna, Austria 
    \and School of Mathematics and Physics, University of Tasmania, Private Bag 37, Hobart, Tasmania 7001, Australia 
    \and NASA Exoplanet Science Institute, Caltech, MS 100-22, 770 South Wilson Avenue, Pasadena, CA 91125, USA 
    \and Perth Observatory, Walnut Road, Bickley, Perth 6076, Australia 
    \and South African Astronomical Observatory, P.O. Box 9 Observatory 7925, South Africa 
    \and Space Telescope Science Institute, 3700 San Martin Drive, Baltimore, MD 21218, USA 
    \and Department of Physics, Institute for Basic Science Research, Chungbuk National University, Chongju 361-763, Korea 
    \and Korea Astronomy and Space Science Institute, 61-1, Whaam-Dong, Youseong-Gu, Daejeon 305-348, Korea 
    \and Bronberg Observatory, Pretoria, South Africa 
    \and Warsaw University Observatory. Al. Ujazdowskie 4, 00-478 Warszawa, Poland 
    \and Universidad de Concepci\'{o}n, Departamento de F\'{\i}sica, Astronomy Group, Casilla 160-C, Concepci\'{o}n, Chile 
    \and Institute of Astronomy, University of Cambridge, Madingley Road, Cambridge CB3 0HA, United Kingdom 
    \and Institute of Information and Mathematical Sciences, Massey University, Private Bag 102-904, North Shore Mail Centre, Auckland, New Zealand 
    \and Solar-Terrestrial Environment Laboratory, Nagoya University, Nagoya, 464-8601, Japan 
    \and Jodrell Bank Centre for Astrophysics, University of Manchester, Manchester M13 9PL, United Kingdom 
    \and Department of Physics, Konan University, Nishiokamoto 8-9-1, Kobe 658-8501, Japan 
    \and Nagano National College of Technology, Nagano 381-8550, Japan 
    \and Department of Physics, University of Auckland, Private Bag 92019, Auckland 1142, New Zealand 
    \and Tokyo Metropolitan College of Industrial Technology, Tokyo 116-0003, Japan 
    \and Department of Physics and Astrophysics, Faculty of Science, Nagoya University, Nagoya 464-8602, Japan 
    \and Mount John Observatory, P.O. Box 56, Lake Tekapo 8770, New Zealand 
    \and Las Cumbres Observatory, 6740B Cortona Dr, suite 102, Goleta, CA 93117, USA 
    \and Astrophysics Research Institute, Liverpool John Moores University, Twelve Quays House, Egerton Wharf, Birkenhead CH41~1LD, United Kingdom 
    \and School of Physics, University of Exeter, Stocker Road, Exeter EX4 4QL, United Kingdom 
    \and Department of Physics, University of Warwick, Coventry, CV4~7AL, United Kingdom 
    \and Department of Physics, Broida Hall, University of California, Santa Barbara CA 93106-9530, USA 
    \and LATT, Universit\'{e} de Toulouse, CNRS, 14 avenue Edouard Belin, 31400 Toulouse, France 
    \and Niels Bohr Institute, K{\o}benhavns Universitet, Juliane Maries Vej 30, DK-2100 K{\o}benhavn, Denmark 
    \and University Observatory Munich, Scheinerstrasse 1, 81679 M\"unchen, Germany 
    \and Max Planck Institute for Extraterrestrial Physics, Giessenbachstrasse, 85748 Garching, Germany 
    \and Institute of Theoretical Physics, Charles University, V Hole\v{s}ovi\v{c}k\'{a}ch 2, 18000 Prague, Czech Republic 
    \and Bellatrix Observatory, Via Madonna de Loco 47, 03023 Ceccano, Italy 
    \and Dipartimento di Fisica, Universita' di Salerno and INFN, sez. di Napoli, Italy 
% ...
}
  \date{ Received / Accepted }
  \abstract
      {Gravitational microlensing is not only a successful tool for discovering
distant exoplanets, but it also enables characterization of the lens and source
stars involved in the lensing event.}
      {In high magnification events, the lens caustic may cross over the source
disk, which allows a determination of the angular size of the source and 
additionally a measurement of its limb darkening.}
      {When such extended-source effects appear close to maximum magnification,
the resulting light curve differs from the characteristic Paczy\'nski 
point-source curve. The exact shape of the light curve close to the peak 
depends on the limb darkening of the source. Dense photometric coverage permits
measurement of the respective limb-darkening coefficients.}
      {In the case of microlensing event OGLE~2008-BLG-290, the K giant source 
star reached a peak magnification of about 100. Thirteen different telescopes 
have covered this event in eight different photometric bands. Subsequent 
light-curve analysis yielded measurements of linear limb-darkening coefficients
of the source in six photometric bands. The best-measured coefficients lead to
an estimate of the source effective temperature of about $4700^{+100}_{-200}$~K.
However, the photometric estimate from colour-magnitude diagrams favours a 
cooler temperature of $4200 \pm 100$~K.}
      {As the limb-darkening measurements, at least in the CTIO/SMARTS2 $\Vs-$ 
and $\Is-$ bands, are among the most accurate obtained, the above disagreement 
needs to be understood. A solution is proposed, which may apply to previous 
events where such a discrepancy also appeared.}
\maketitle

\section{Introduction} \label{sec:intro}

Astrophysical opportunities for measuring stellar limb darkening are very 
scarce. Moreover, they are often limited to very specific types of stars. Up to
recently, most measurements had been based on analyses of eclipsing-binary 
light curves \citep{Popper1984}, requiring large amounts of high-precision 
photometric data but only rarely yielding useful constraints on the recovered 
limb-darkening parameters \citep[e.g.][]{PE1981,Southworth2005}.

Resolving stellar surfaces by interferometry is a promising technique, limited 
nevertheless at present to nearby giants and supergiants. For most of the 
sufficiently well-resolved stars it has been shown that a fixed 
model-atmosphere-based limb-darkening profile agreed with the observations 
better than a uniform or fully limb-darkened stellar disk. Only a limited 
number of cases have been used to actually measure limb darkening from 
interferometric data 
\citep[e.g.,][]{Burns1997,Perrin2004,Aufdenberg2006,Wittkowski2006}. The same 
is true for observations of lunar occultation of stars, which yield 
limb-darkening measurements only rarely \citep{Richichi1990}.

An alternative method which permits accurate limb-darkening measurement 
practically independent of stellar type is based on gravitational microlensing.
In a stellar gravitational microlensing event the flux from an observed source 
star is temporarily amplified by the gravitational field of another object 
passing in the foreground and acting as a gravitational lens 
\citep{Paczynski1996}. The lensing object may be typically a single star, a 
binary star, or a star with one or several planetary companions. A particularly
interesting situation occurs in so-called caustic-crossing events, in which the
caustic of the lens directly crosses the projected disk of the source star, 
typically on a timescale of a fraction of a day to a few days. The very high 
angular resolution provided by the caustic presents a unique opportunity to 
measure the source star's limb darkening from photometric observations or the 
centre-to-limb variation of spectral features from spectroscopic observations 
of the crossing \citep[e.g.][]{Witt1995,HSL2000,Heyrovsky2003}.

The first well-documented source-resolving caustic-crossing microlensing event 
was MACHO Alert 95-30 \citep{Alcock1997}, in which an M4~III giant source star 
was scanned by the point-like caustic of a single lens. The observations 
permitted measurement of the source size and detection of its limb darkening. 
Precise measurement of the limb darkening was not possible due to low-amplitude
variability of the source \citep{Heyrovsky2003}. The first limb-darkening
measurement came from the analysis of binary-lens event MACHO~97-BLG-28
\citep{Albrow1999b}, producing $V-$ and $I-$band coefficients of the linear and
square-root limb-darkening laws for the K2~III giant source. By the time of 
writing, six further binary-lens caustic-crossing events yielded limb-darkening
measurements of their source stars: MACHO~97-BLG-41 \citep{Albrow2000b}, 
MACHO~98-SMC-1 \citep{Afonso2000}, OGLE~1999-BUL-23 \citep{Albrow2001a}, 
EROS~2000-BLG-5 \citep{An2002,Fields2003}, MOA~2002-BLG-33 \citep{Abe2003}, and
OGLE~2002-BLG-069 \citep{Cassan2004,Kubas2005}.

Limb-darkening measurements are also available from four single-lens 
caustic-crossing events: OGLE~2003-BLG-262 \citep{Yoo2004}, OGLE~2003-BLG-238 
\citep{Jiang2004}, OGLE~2004-BLG-254 \citep{Cassan2006,Heyrovsky2008}, and 
OGLE~2004-BLG-482 \citep{Zub2009}. In addition, there are a comparable number 
of events that required the inclusion of limb darkening in their analysis (such
as MACHO Alert 95-30), but for which a sufficiently accurate limb-darkening 
measurement was not (and often could not be) performed. These include two 
further single-lens events \citep{Yee2009,Batista2009}, as well as five of the 
eight published planetary microlensing events, in which a star with a planet
acted as the lens \citep{Bond2004,Dong2009b,Dong2009a,Beaulieu2006,Janczak2010}.

In this paper we report the results of the analysis of OGLE~2008-BLG-290, a 
microlensing event in which the caustic of a single lens crossed the disk of a 
K giant in the Galactic Bulge. In \Sec{sec:data} we describe the obtained data 
and their reduction. We discuss the properties of the source star in detail in 
\Sec{sec:source}. The data modelling in \Sec{sec:model} is followed in 
\Sec{sec:LDmeasure} by a description of the limb-darkening measurements in six 
photometric bands and their comparison with model atmospheres. We devote 
\Sec{sec:lens} to the study of the properties of the lens. We explore the
possibility of deviations caused by a potential planetary companion to the lens
in \Sec{sec:planet} and consider other potential systematics in 
\Sec{sec:systematics}. The main results are summarized in \Sec{sec:discussion}.

\section{Data sets: observations and data reductions} \label{sec:data}

The OGLE-III Early Warning System (EWS) \citep{Udalski2003a} alerted the Bulge
event OGLE~2008--BLG--290 ($\alpha=$~17h58m29.3s,
$\delta=-27\degr 59\arcmin 22\farcs$ (J2000.0) and $l=2.256\degr$,
$b=-1.951\degr$) on May 15, 2008, from observations carried out with
the 1.3 m Warsaw Telescope at the Las Campanas Observatory (Chile).
Independently, the MOA-II 1.8m telescope at Mount John Observatory (New Zealand)
discovered the same event under name MOA~2008--BLG--241 on May 31, 2008.

A few days later, it was clear that this event had the potential to become one
of the rare high magnification events discovered each year, and follow-up
observations were undertaken on PLANET, $\mu$FUN, RoboNet/LCOGT and
MiNDSTEp telescopes available at that time. In total, 13 telescopes covered the
event in different photometric bands: OGLE 1.3m ($I-$band), MOA-II 1.8m (wide
MOA-red band), SAAO 1.0m at Sutherland (South Africa) ($V-$ and $I-$bands),
Canopus 1.0m at Hobart (Australia) ($I-$band), Perth/Lowell 0.6m at Bickley
(Australia) ($I-$band), CTIO 1.3m at Cerro Tololo (Chile) ($\Vs-$, $\Is-$ and
$H-$bands), LOAO 1.0m at Mount Lemmon (Arizona) ($I-$band), Bronberg 0.36m at
Pretoria (South Africa) (unfiltered), Skinakas 1.3m at Mount Ida (Greece)
($I-$band), Faulkes North 2.0m at Haleakala (Hawaii) (Bessell $R-$band),
Faulkes South 2.0m at Siding Spring (Australia) (Bessell $R-$band), Liverpool
2.0m at La Palma (Spain) (SDSS $r-$band), and Danish 1.5m at La Silla (Chile)
($I-$band). The CTIO/SMARTS2 bandpasses are close to standard Bessell
$V$ and $I$, but sufficiently distinct to be treated separately
\citep{Gould2010}. We mark them here by $\Vs$ and $\Is$.

Thanks to the public availability of data from the different groups, real-time
modelling efforts then showed that the light curve was deviating from a normal
Paczy\'nski curve \citep{Paczynski1986}, exhibiting evidence of extended-source
effects. The event peaked on that same night (June 15). Because such events are
reasonably sensitive to Jupiter-mass planets, as recently shown by
\citet{Dong2009b}, amateur telescopes from the $\mu$FUN network were
alerted, resulting in an excellent coverage of the peak region from 9 different
telescopes.

Data reduction has been done using both PSF photometry based on a customized
DoPhot package and image subtraction. $\mu$FUN telescope images were first
reduced using DoPhot. RoboNet/LCOGT data were reduced using an automatic image
subtraction package, and have been re-reduced off-line. PLANET telescopes also
use image subtraction: at telescope an on-line version called WISIS, based on
Alard's ISIS package, then off-line version 3.0 of pySIS \citep{Albrow2009},
based on a numerical kernel \citep{Bramich2008}. SAAO $I$ photometry obtained
with pySIS has been checked independently using a DIA package. However,
even then the SAAO $I$ data were found to exhibit correlated noise at the peak.
Its effect was found to be strong enough to skew the recovered lensing
parameters (see \Sec{sec:model}). For this reason we decided to omit the light
curve from the final event analysis, even though we include it when testing for
a potential planetary companion in \Sec{sec:planet}. SAAO $V$ images were
taken well after the light curve peak and therefore do not bring constraints on
the event parameters. Bronberg images were re-reduced using the same
DIA package. Looking at stars of similar colours as the target, a clear
airmass effect is generally detected in Bronberg data because there
is no filter on the camera. However, in the case of this event, only a
clear trend with time was present in the data, instead of the expected airmass
trend. As the origin of this trend is unexplained, we could not correct for it
and as a result we did not use Bronberg data in the following analysis. CTIO
$\Is$ and $\Vs$ and LOAO $I$ images were re-reduced off-line using
pySIS~3.0, with a slight improvement over DoPhot. CTIO $H$ images were of poor
quality and thus discarded. Danish images were reduced using pySIS~3.0; they
were also reduced with the DIAPL package from Pych \& Wo\'zniak
\citep{Wozniak2000}, with very similar results. Finally, OGLE images were
re-reduced with the OGLE pipeline, but with a better resolution reference image
and a correctly adjusted centroid.

The final data set before rejection of outliers contains 4389 data
points from 11 different telescopes (OGLE $I$: 1065, CTIO $\Is$: 252,
LOAO $I$: 159, CTIO $\Vs$: 12, MOA-red: 2555, Canopus $I$: 44,
Perth $I$: 13, Skinakas $I$: 28, Faulkes North $R$: 12, Faulkes South $R$: 27,
Liverpool $r$: 59, Danish $I$: 163).

\begin{figure}[ht]
    \includegraphics[width=8.5cm]{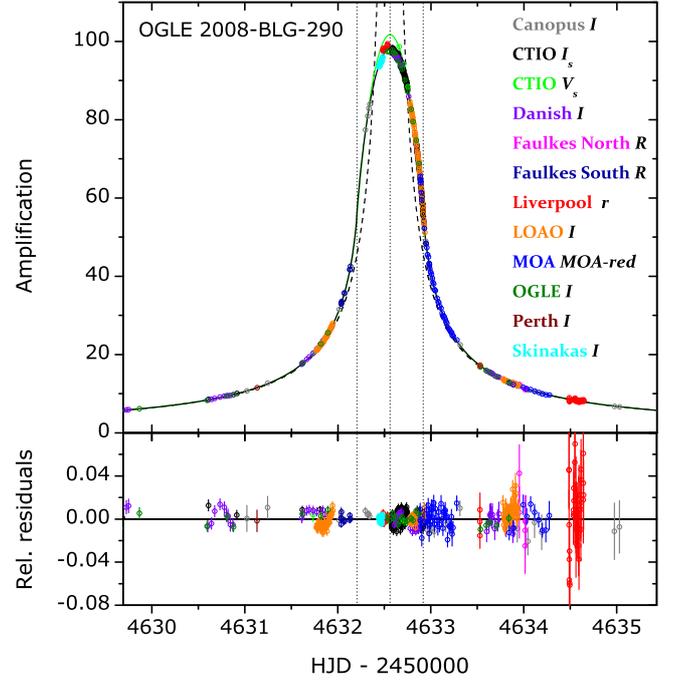}
    \caption{Upper panel: photometry of microlensing event
    OGLE~2008--BLG--290 in terms of source-flux amplification, as observed by
    OGLE (olive), MOA (blue), two PLANET sites (Canopus: grey;
    Perth: wine), three $\mu$FUN sites (CTIO 1.3m: $\Is$: black, $\Vs$: green;
    LOAO: orange; Skinakas: cyan), three RoboNet/LCOGT sites (Faulkes
    North: magenta; Faulkes South: royal blue; Liverpool: red),
    and one MiNDSTEp site (Danish: violet). The amplification is determined
    using the limb-darkened source fit with parameters given in
    \Tab{tab:fitparameters}. Plotted lines correspond to the $I$-band (solid
    black) and $\Vs$-band (solid green) light curve solutions, plus a
    point-source light curve with same impact parameter (dashed black) for
    comparison. Lower panel: relative flux residuals of the combined
    data set (the full time span of the analyzed 2008 data runs from $t$=4502
    to 4769). The vertical dotted lines show the times of source limb entry,
    closest approach, and limb exit.}
    \label{fig:lc_all}
\end{figure}

\begin{figure}[ht]
    \includegraphics[width=8.5cm]{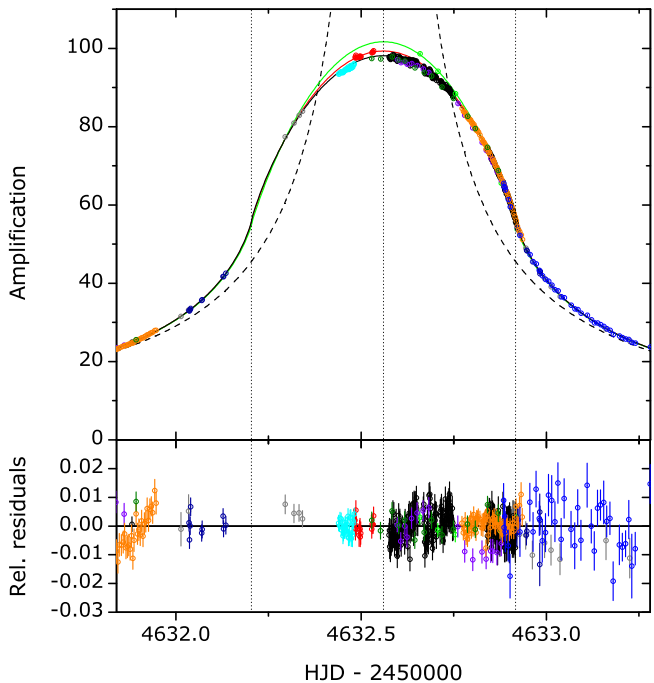}
    \caption{Zoom of the event peak (June 15, 2008) from \Fig{fig:lc_all}.
    Amplification is plotted in the upper panel; relative residuals in the
    lower panel. Colours and lines have the same meaning as in
    \Fig{fig:lc_all}, with the added SDSS $r$-band light curve (solid red).}
    \label{fig:lc_zoom}
\end{figure}

The shape of the light curve depends on the limb darkening of the source which
is different for each photometric band. In the following two figures we plot by
solid lines only those that are distinguishable at the peak: the $I-$ and
$\Vs-$ band light curves in \Fig{fig:lc_all}, and the $I-$,
$\Vs-$ and $r$-band light curves in \Fig{fig:lc_zoom}. The curves
correspond to the best-fit limb-darkened extended-source model described
further and specified in \Tab{tab:fitparameters}. The residuals in the lower
panels are computed and displayed for their respective light-curve solutions.
Dashed lines in these figures correspond to a point-source light curve with the
same timescale and impact parameter.

\section{Source properties} \label{sec:source}

\subsection{Near-infrared Colour-Magnitude Diagram}

The distance to the source and the amount of reddening along the line of sight
are uncertainties which always affect the final determination of the properties
of the lens - source system. We want to dedicate a short discussion to these
issues to justify our adopted values and associated uncertainties and to serve
as a reference in our future papers.

Due to the geometry of the Galactic Bulge with a bar embedded in it, the
galactic coordinates of the target give an estimate of the relative position of
the source with respect to the Galactic Centre, if we assume that the source is
at the same distance as the majority of the stars in the field. The Galactic
Centre distance itself is adopted from \citet{Eisenhauer2005} as
$D_{\rm GC} = 7.62 \pm 0.32$ kpc or $\mu_{\rm GC} = 14.41 \pm 0.09$.
\citet{Rattenbury2007} give the relative positions of the OGLE-II fields with
respect to the field BUL\_SC45, which contains Baade's Window
($l=1.00^{\circ}$, $b=-3.88^{\circ}$). As shown by \citet{Paczynski1998}, it is
safe to assume that the mean distance of stars seen in Baade's Window is
similar to the Galactic Centre distance. Our target's position lies outside of
any OGLE-II fields, but close to field BUL\_SC20, and this field is at about
the same distance as BUL\_SC45. We therefore adopt the Galactic Centre distance
modulus as the source distance modulus, but with an increased uncertainty due
to the additional involved assumptions, namely $\mu = 14.4 \pm 0.2$.

There are several estimates of the reddening in the $\Ks-$band at about the
position of our target. They range from $\Aks = 0.28$ \citep{Schultheis1999} to
$\Aks = 0.46$ \citep{Dutra2003}. Part of the disagreement may come from the
patchiness of dust structure, but unfortunately different assumptions about the
reddening law also play a role.

Our own estimate is based on IRSF/Sirius photometry of a 7.7' $\times$ 7.7'
field containing our target. We use isochrones from \citet{Bonatto2004} based
on Padova group models, but directly calibrated for the 2MASS bandpasses. The
IRSF/Sirius photometric system, similar to the MKO system \citep{Tokunaga2002},
is close to the 2MASS system, but we calibrated the photometry using 2MASS
stars in the same field to ensure coherence.

\begin{figure}[ht]
  \centering
    \includegraphics[width=9cm]{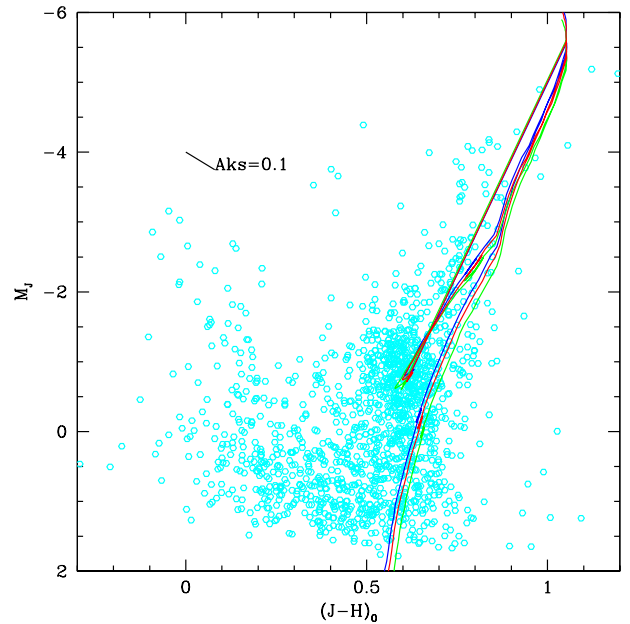}
    \caption{Colour-magnitude diagram in $J$ and $H$ from IRSF/Sirius stars in
     a 2.25' $\times$ 2.25' field around OGLE~2008--BLG-290 assuming a distance
     modulus of 14.4, and extinction coefficients of $\Aj=1.10$, $\Ah=0.75$ and
     $\Aks=0.44$. Superimposed are isochrones from \citet{Bonatto2004} for
     three ages (10, 12.6, 17.8 Gyr, from left to right in colour) assuming
     solar metallicity.}
    \label{fig:cmd_IRSF}
\end{figure}

We restrict the fitting region to 300 pixels around the target (2.25' $\times$
2.25') to avoid differential extinction. This is large enough to form
well-defined colour-magnitude diagrams (hereafter CMD), where the red
giant clump (hereafter RC) is easily identified. This is not the case
when using only 2MASS, because the brighter limiting magnitude cuts off part of
the clump. An example of such a CMD is displayed in \Fig{fig:cmd_IRSF}.

Although the position of the clump could in principle give an estimate of its
distance, in practice variations in age and metallicity do not allow an
accurate determination. Adopting a 10 Gyr isochrone and a solar metallicity, we
get the following estimates of the near-infrared extinction and reddening law:

\begin{equation}
\Aks = 0.44
\end{equation}
\begin{equation}
\Aj = 2.5 \times \Aks
\label{eq:RL_JK}
\end{equation}
\begin{equation}
\Ah = 1.7 \times \Aks
\label{eq:RL_HK}
\end{equation}

Note that the corresponding reddening law in $J$ differs from what
\citet{Nishiyama2006} found based on the IRSF/Sirius Galactic Bulge survey
($\Aj = 3.0 \times \Aks$), obtained with the same telescope and
instrumentation. This possibly reflects variations of the reddening law in
different directions of the Galactic Bulge.

The target, which appears as a double star in the OGLE finding chart, is well
separated by PSF photometry of the IRSF/Sirius image. The accurate coordinates
and magnitudes of the two components are given in \Tab{tab:components},
together with the corresponding values for the single object in the 2MASS
catalogue.

\begin{table*}[ht]
  \caption{Coordinates and magnitudes of the two stars close to the target
   position and the 2MASS blend.}
  \label{tab:components}
  \centering
  \begin{tabular}{lllccc}
    \hline\hline
      Designation & $\alpha$ (2000) & $\delta$ (2000) & J & H & $\Ks$ \\
    \hline
      & \\
Brighter Northeastern component & 17:58:29.362 & -27:59:21.45 & $14.291 \pm 0.040$ & $13.369 \pm 0.026$ & $13.042 \pm 0.051$ \\
Fainter Southwestern component & 17:58:29.296 & -27:59:22.00 & $14.947 \pm 0.054$ & $13.916 \pm 0.038$ & $13.447 \pm 0.023$ \\
2MASS~17582933-2759215 & 17:58:29.33 & -27:59:21.5 & $13.898 \pm 0.034$ & $12.936 \pm 0.043$ & $12.506 \pm 0.034$ \\
      & \\
    \hline
  \end{tabular}
\end{table*}

Although the 2MASS flags do not indicate any blending, the coordinates and
magnitudes correspond well to the blend of the two stars. The microlensed
target is the fainter southwestern component and, after extinction correction
and conversion to the Bessell \& Brett photometric system
\citep{Bessell1988},
it has $\Ko = 13.05$ and typical colours of a K4 giant star ($\JKo=0.87$,
$\HKo=0.13$). Such a star is predicted to have colours of $\VIo=1.50$ and
$\VKo=3.26$, so we expect $\Vo=16.3$ and $\Io=14.8$.


Estimating the uncertainties of the measured near-infrared extinctions
is not straightforward, because the present method mixes hypotheses about
distance, age, metallicity with actual measurements by isochrone fitting. In
the next paper of this series \citep{Fouque2010} on MOA~2009-BLG-411, a refined
method will be introduced, using reddening law, clump absolute magnitudes and
distance hypothesis to simultaneously fit isochrones to near-infrared and
visible CMDs and derive a coherent source size. We applied this new technique to
the present data set to estimate the uncertainties in our measured extinctions.
We found differences of 0.06, 0.13 and 0.05 in $\Aj$, $\Ah$ and $\Aks$,
respectively. We therefore adopt an uncertainty of 0.1 in the extinction
measurement for each near-infrared band.

\subsection{Visible CMD} \label{sec:viscmd}

\begin{figure}[ht]
  \centering
    \includegraphics[width=8cm]{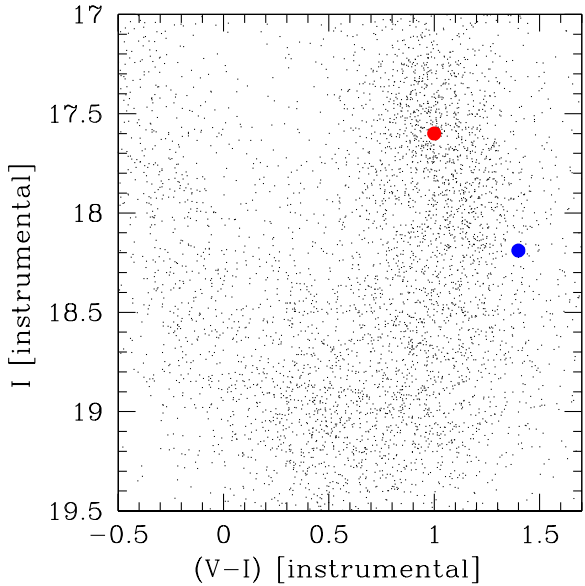}
    \caption{Colour-magnitude diagram in $\Is$ and $\Vs$ from CTIO 1.3m
     telescope. The blue disk shows the source position, while the red one
     marks the centroid position of the red giant clump.}
    \label{fig:cmd_ctio}
\end{figure}

A colour-magnitude diagram allows an estimate of the dereddened magnitude and
colour of the source, by comparison with the observed position of the red giant
clump, if one assumes that both suffer the same amount of extinction. The
CTIO 1.3m telescope obtained the uncalibrated colour-magnitude diagram
displayed in \Fig{fig:cmd_ctio} (instrumental magnitudes), from which we read a
magnitude shift in I of $0.56 \pm 0.1$ for the source compared to the RC, and a
colour shift of $0.37 \pm 0.1$.

For the mean absolute magnitude of the clump, we adopt the Hipparcos value as
given by \citet{Stanek1998}, $\Mi=-0.23 \pm 0.03$ and for the mean colour
${\VIo}^{\rm RC}=1.00 \pm 0.05$. With our adopted distance to the field,
$\mu=14.4 \pm 0.2$, this gives $\Io^{\rm RC}=14.17 \pm 0.2$. So our source is
predicted to have $\Io^{\rm S}=14.73 \pm 0.23$ and ${\VIo}^{\rm S}=1.37 \pm 0.11$,
in good agreement with the values obtained before from the near-infrared CMD.

From magnitude and colour, and using the recent revision of surface-brightness
-- colour relations (hereafter SBC) in $I$, $\VI$ published by
\citet{Kervella2008}, we get an estimate of the angular source radius $\ThS$ in
$\muas$ of:

\begin{equation}
\log \ThS = -0.2 \Io + 0.4895 \VIo - 0.0657 \VIo^2 + 3.198
\label{eq:SBVI}
\end{equation}

The uncertainty of this estimate is 0.0238, so adding quadratically the
uncertainty in magnitude and colour gives $\ThS = 6.3 \pm 1.1 \, \muas$.

However, this source is probably a giant star, and these SBC relations have
been calibrated for dwarfs and sub-giants. In order to check if this introduces
a systematic uncertainty in our measurement of the angular radius, we use the
\citet{Groenewegen2004} SBC relation specifically designed for giants:

\begin{equation}
\log \ThS = -0.2 \Ko + 0.045 \VKo + 3.283
\label{eq:SBVK}
\end{equation}

The uncertainty of this estimate is similar at 0.024, so adding quadratically
the uncertainty in the magnitude $\Ko=13.05 \pm 0.06$ and estimated colour
$\VKo=3.1 \pm 0.2$ gives $\ThS = 6.5 \pm 0.4 \,\muas$, in agreement with the
previous estimate. At the adopted source distance, this translates into a
linear radius of $\RS = 10 \ \rsun$, typical of a K giant. Please note that in
the estimate using the visible CMD, the angular radius depends on the adopted
source distance, while the linear radius does not. Conversely, in the
near-infrared method, the angular radius depends very little on the adopted
source distance while the linear radius does. The agreement between both values
shows that our adopted source distance is correct.

Using the dereddened colours and, for instance, \citet{Houdashelt2000}
tables, we estimate the effective temperature of the source star to be
about 4200~K, corresponding to a K3 giant. In order to estimate an
uncertainty of this value, we repeated the whole procedure adopting plausible
different values for the clump distance, its colour and absolute magnitude,
and measured the shift in colour and magnitude of the source with respect to
the centroid of the clump using the OGLE-III photometric catalogue. The
difference in colours that we get using these new values corresponds to a
difference in temperature of about 100~K. We therefore adopt $4200 \pm 100$~K
as our estimate of the photometric temperature of the source star.

Using this value, we looked at \citet{Marigo2008} isochrones for a model star
with similar characteristics to ours. Two routes are used to derive the star's
luminosity: on one hand, from $\Ko = 13.05$, distance modulus (14.4)
and model bolometric correction in K (2.36), we get $\Mbol = 1.01$ and
$\logl = 1.48$; on another hand, we use $\teff$ and our estimate of the
star's radius, to get $\logl = 1.45$. We note that different modelers
still use different values of the bolometric magnitude of the Sun
(4.72 for \citet{Houdashelt2000} vs. 4.77 for \citet{Marigo2008}). We
find that an old star (12.7 Gyr), slightly metal-rich (Z=0.03) gives a red
giant of 1 solar mass and $\logg = 2.5$, with such characteristics.

\section{Data modelling: noise model} \label{sec:model}

From the original data set, we remove data points in OGLE and MOA older than
date 4490 or newer than 4770. This corresponds to selecting the whole 2008
observing season.
We verified that this does not change the resulting fit parameters. The
reason for this cut is two-fold: the planetary deviation search is very
demanding in terms of CPU time, so reducing the number of useful points helps;
moreover, the number of data points in the baseline before 4490 is quite large,
and any slight error in the photometric error estimate may bias the fit.

We then proceed to remove outliers and rescale photometric error bars in a
consistent way: the rescaling factor is computed by forcing $\chidof
\simeq 1$ for each telescope data set independently, and after rescaling,
any point lying at more than $3 \  \sigma$ is removed. This slightly changes
the estimate of $\chisq$ and thus the rescaling factor. After a few iterations,
the process converges. Care was taken not to remove any potential planet-caused
outliers. The number of outliers varies between 1 and 6 per telescope.

We return briefly to the SAAO $I-$band light curve. Inspecting the
residuals of the microlensing fit, we find they are mutually correlated rather
than randomly distributed, as mentioned earlier in \Sec{sec:data}. Near the
event peak clusters of points with positive residuals alternate with
negative-residual clusters. Despite its small amplitude, the effect is strong
enough to influence the recovered lensing parameters. Adding the SAAO light
curve to the twelve other curves results in a lower impact parameter and
shorter source-radius crossing time, both of which in turn influence the
limb-darkening measurement. The small-amplitude abrupt shifts in the SAAO
residuals during the crossing are most probably a systematic artifact in the
data (see also \Sec{sec:planet} and \Sec{sec:systematics}). We therefore
decided not to include this light curve in the subsequent event analysis.

In the case of MOA data, several measurements close to the peak were saturated
and therefore were eliminated. For this particular telescope, 51 data points
were removed a priori or as outliers. For OGLE data, we quadratically added
0.003 mag to all photometric errors, in order to avoid too small errors at the
peak, which correspond to Poisson noise at these bright magnitudes, but do not
reflect systematic errors.

The final total number of fitted data points amounts to 2049 in 12
light curves observed in 6 different photometric bands (see \Fig{fig:lc_all}).

\section{Limb darkening of the source star} \label{sec:LDmeasure}

\subsection{Limb-darkening measurement} \label{sec:LDM}

As expected, a single-lens point-source model results in a high
$\chidof \sim 40$, which obviously rejects this simple model. A uniform
extended-source model provides us with a first working set of
parameters; however, the fit still has a high $\chidof \sim 3.6.$ At
this stage the residuals of the fit show a symmetric pattern around
the peak of the light curve, the signature of limb darkening. We thus add
linear limb darkening to the source model. This is described here by
\begin{equation}
I(r)\propto 1 - u \left( 1-\sqrt{1-r^2}\right),
\label{eq:LLD}
\end{equation}
where $r$ is a radial coordinate running from 0 at disk centre to 1 at limb,
and $u$ is the linear limb-darkening coefficient
\citep[hereafter LLDC, e.g.,][]{Claret2000}.

The adopted event parametrization involves the following microlensing
quantities: $t_{\circ}$ (time of closest approach), $u_{\circ}$ (impact parameter
in units of the Einstein ring radius $\ThE$), $\tE$ (crossing time of $\ThE$),
$\rhoS$ (source radius in units of $\ThE$) and two annual parallax
parameters $\piE$ and $\psi$ (see \Sec{sec:lens}) common to all
data sets; plus light-curve-specific parameters: baseline and
blending fluxes, as well as the LLDC $u$ for the respective photometric band.
The main parameters of the set best fitting the data ($\chidof = 1.15$)
and their errors are given in \Tab{tab:fitparameters}. The
limb-darkening coefficients correspond to the involved six distinct photometric
bands: CTIO/SMARTS2 $\Vs$, SDSS $r$, $R$, MOA-red, CTIO/SMARTS2 $\Is$, and $I$,
in order of increasing effective wavelength.

\begin{table}[ht]
  \caption{Main parameters of the best-fitting point-lens limb-darkened
      source model, using the linear limb-darkening law.}
  \label{tab:fitparameters}
  \centering
  \begin{tabular}{p{3cm}l}
    \hline\hline
      Parameter & Value \& Errors \\
    \hline
      & \\
      ~~~ $\to$ (HJD-2,450,000)\dotfill & $4632.56037\ {\scriptstyle (-0.00027 +0.00019)}$ ~~~\\
      ~~~ $\uo$\dotfill & $0.00276 \ {\scriptstyle (-0.00020 +0.00021)}$ ~~~\\
      ~~~ $\tE$ (days)\dotfill & $16.361 \ {\scriptstyle (-0.077 +0.071)}$ ~~~\\
      ~~~ $\rhoS$\dotfill & $0.02195 \ {\scriptstyle (-0.00011 +0.00010)}$ ~~~\\
      ~~~ $u_{\Vs}$ \dotfill & $0.773 \ {\scriptstyle (-0.011 +0.014)}$ ~~~\\
      ~~~ $u_r$ \dotfill & $0.642 \ {\scriptstyle (-0.058 +0.087)}$~~~\\
      ~~~ $u_R$ \dotfill & $0.617 \ {\scriptstyle (-0.046 +0.071)}$~~~\\
      ~~~ $u_{\rm MOA}$ \dotfill & $0.593 \ {\scriptstyle (-0.037 +0.041)}$~~~\\
      ~~~ $u_{\Is}$ \dotfill & $0.569 \ {\scriptstyle (-0.007 +0.005)}$~~~\\
      ~~~ $u_{I}$ \dotfill & $0.528 \ {\scriptstyle (-0.008 +0.009)}$~~~\\
      ~~~ $\chidof$\dotfill & $2317.7/(2049-34)=1.150$ ~~~\\
      & \\
    \hline
  \end{tabular}
\end{table}

Based on the final model, we find that the source-radius transit time
is $\rhoS \  \tE \simeq 0.36$~d; the time (in terms of $t={\rm HJD}-2,450,000$)
at which the lens starts to transit the source disk $t_{\rm entry}
\simeq 4632.204$~d, while the exit time is $t_{\rm exit} \simeq 4632.917$~d. The
peak magnification achieved in the $I-$band is $\sim 97$, in the $\Vs-$band
$\sim 102$. The source $I-$band magnitude in the OGLE-III photometric database
is 17.26, so the source star temporarily brightened to magnitude 12.3 due to
the lens passing in the foreground.

\begin{figure}[ht]
 \centering
  \includegraphics[width=9cm]{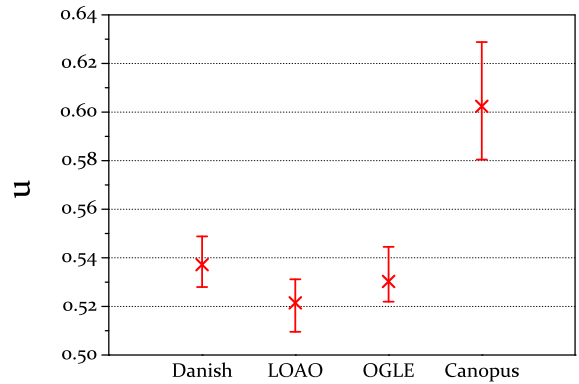}
  \caption{Linear limb-darkening coefficients in the $I-$band for 4 different
   telescopes when fitting the limb darkening independently for each data set.
   We determine the $I-$band measurement as the weighted mean of these
   values, $u_I=0.539\pm0.033$.}
  \label{fig:Iband_LLDC}
\end{figure}

Individual telescopes constrain the measured limb darkening unequally,
because of their different coverage of the caustic-crossing peak. As a result,
Perth, Skinakas and Faulkes-North do not contribute, Faulkes-South very weakly
(closest points to the peak are just outside the limb) and Liverpool only
weakly. Hence, the obtained limb darkening in the $R-$band (Faulkes telescopes)
and the $r$-band (Liverpool) is poorly constrained. In the $\Vs-$ and $\Is-$
bands, the LLDCs rely on the respective CTIO light curves. With the MOA-red
band result determined by MOA data, only the $I-$band result is constrained by
4 different telescope data sets (Danish, LOAO, OGLE, Canopus).

In order to test for consistency we fit the event data again, using
common lensing parameters but leaving the limb darkening independent for each
of the 12 data sets. Checking the $I-$band LLDCs shown in \Fig{fig:Iband_LLDC},
we find that even though the Canopus value is consistent with the others at the
3-$\sigma$ level, the scatter of the values exceeds individual errors as well
as the error obtained from the straightforward fit in \Tab{tab:fitparameters}.
In the light of this finding, the small error on $u_{I}$ in
\Tab{tab:fitparameters} is not really the result of an excellent agreement
between individual light curves. Rather, it is a statistical artifact obtained
by the competing higher-$u$-value Canopus and the lower-$u$-value-favouring
others. Possible causes of this discrepancy include slightly different filters,
detector or atmosphere response, light curve coverage, secondary amendments of
the source or lens models, but also noise models for each telescope. To get a
more realistic result under these circumstances, we determine the $I$-band LLDC
from the values in \Fig{fig:Iband_LLDC}. The weighted mean of the individual
values with the statistical error given by their scatter yields
$u_I=0.539\pm0.033$.

\subsection{Comparison with model atmospheres} \label{sec:comparison}

We obtain the response functions of the individual telescopes by
combining their filter transmission curves and their CCD quantum efficiency
curves. We then use the response functions to compute theoretical values of
the LLDCs, based on Kurucz's ATLAS9 atmosphere models
\citep{Kurucz1993a,Kurucz1993b,Kurucz1994}. To get the LLDC for a
particular theoretical limb-darkening profile we use the radially integrated
fit method described in \citet{Heyrovsky2007} rather than the 11-point fit from
\citet{Claret2000}, for reasons described in the former reference. We compute
the LLDC values for a sub-grid of Kurucz's model grid based on the source
properties inferred in \Sec{sec:source}, with effective temperature $\teff$
ranging from 4000~K to 5000~K, metallicity [Fe/H] from 0 to $+0.3$, surface
gravity $\logg$ from 2.0 to 3.0, and microturbulent velocity fixed at
$v_t=2$~km~s$^{-1}$.

\begin{figure}[ht]
  \includegraphics[width=9.5cm]{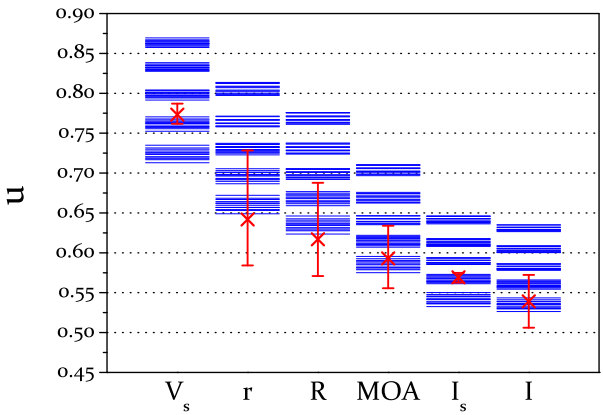}
  \caption{Linear limb-darkening coefficients of the K-giant source star
in CTIO/SMARTS2 $\Vs-$, SDSS $r-$, $R-$, MOA-red, CTIO/SMARTS2 $\Is-$,
and $I-$bands, compared to theoretical model atmosphere predictions (blue) for
different effective temperatures (five broad bands in 250~K steps from 4000~K
at top to 5000~K at bottom), different metallicities (from [Fe/H]=0.3 at top to
0.0 at bottom), and different surface gravities (finest structure with
$\logg=2.0$, 2.5, 3.0) at each temperature.}
  \label{fig:VRMIbands_LLDC}
\end{figure}

The comparison of the measured LLDC in the 6 bands to theoretical model
predictions is presented in \Fig{fig:VRMIbands_LLDC}. Judging first the
agreement between the bands, we point out that the obtained values decrease
with increasing effective wavelength, in agreement with theoretical
expectations. Second, we note that within each band the measured values point
to a similar effective temperature of the source, with all measurements being
mutually compatible within their error bars. This leads to an LLDC-based
initial estimate of the effective temperature of the source star, with
the most accurate $\Vs$ and $\Is$ measurements indicating 4750~K. The value is
in rather poor agreement with the photometric estimate 4200~K derived in
\Sec{sec:viscmd}. We note that the theoretical \citet{Claret2000} coefficients
would lead to an even larger discrepancy, with the $I-$band measurement
indicating an effective temperature higher than 5500~K.

Comparing measured LLDCs with corresponding values computed from model
atmospheres has its potential pitfalls \citep{Heyrovsky2003,Heyrovsky2007}.
Simple analytical limb-darkening laws, and the linear law in particular, often
do not describe theoretical limb-darkening profiles sufficiently accurately.
Such analytical approximations do not even conserve the flux of the
approximated profile with the required precision. To avoid this problem at
least partly, we compare our measured LLD profiles directly with the
theoretical profiles of the Kurucz models rather than with their
linear limb-darkening approximations.

\begin{figure*}[ht]
 \centering
  \includegraphics[width=9cm]{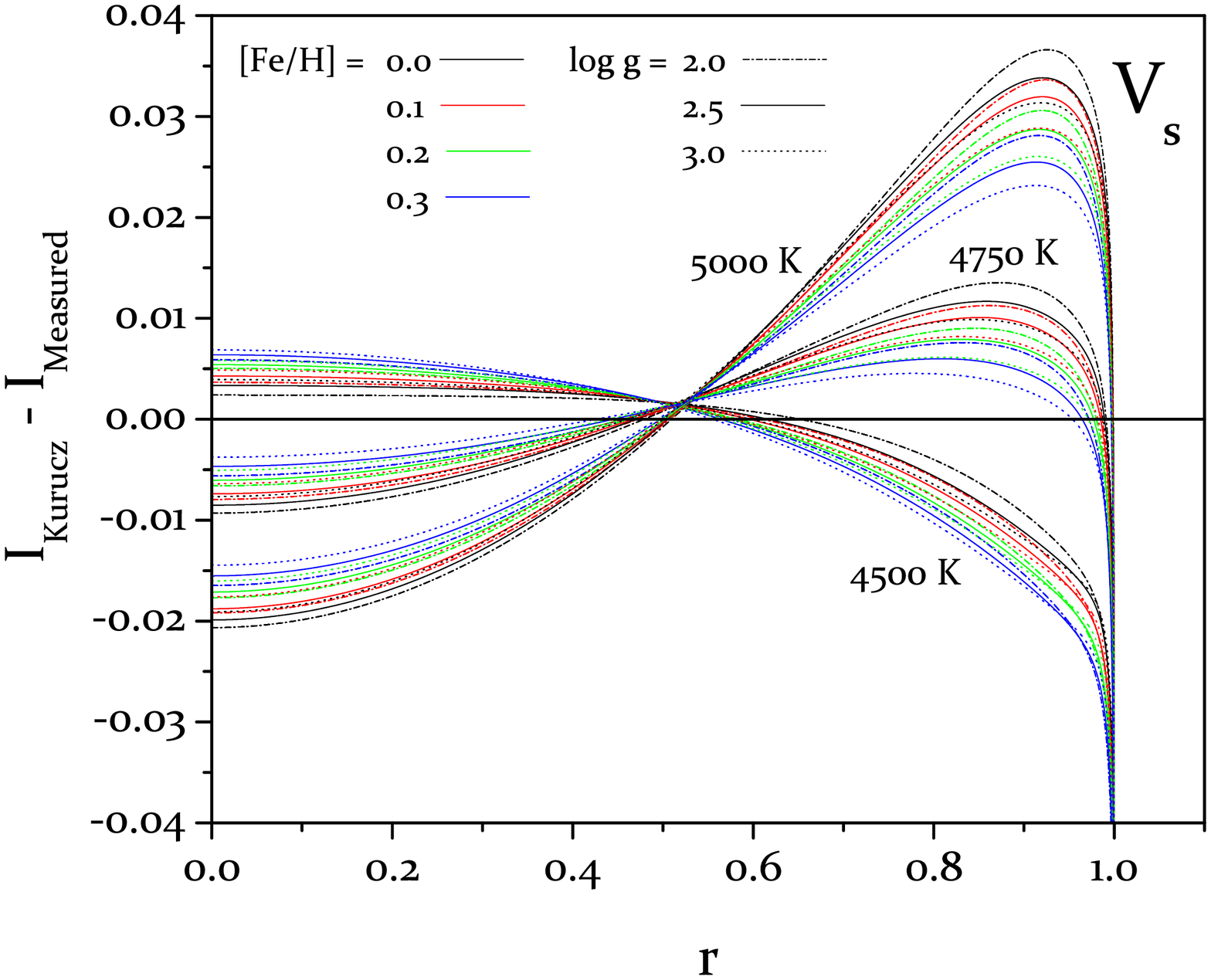}
  \includegraphics[width=9cm]{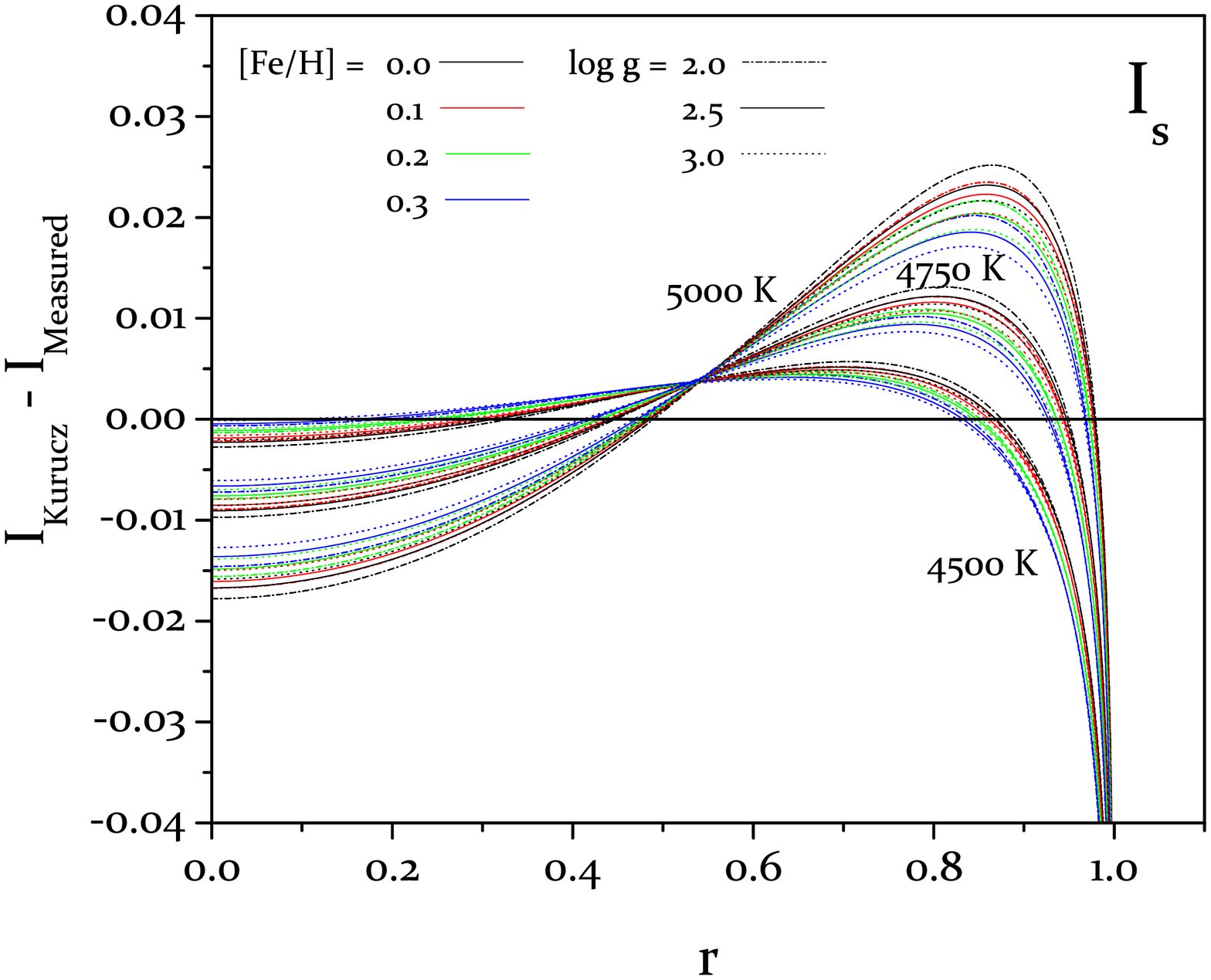}
  \caption{Comparison of the measured $\Vs-$band (left panel) and
  $\Is-$band (right panel) limb-darkening profiles with corresponding
  theoretical limb-darkening profiles from Kurucz's ATLAS9 model-atmosphere
  grid: absolute difference $I_{\rm Kurucz}-I_{\rm Measured}$ as a function of
  radial position $r$ on the stellar disk. $I_{\rm Measured}$ is normalized to
  unit r.m.s. intensity; each theoretical profile $I_{\rm Kurucz}$ is scaled
  vertically to minimize the r.m.s. difference. The three broad bands
  formed by the curves correspond to three effective temperatures $\teff$,
  colours indicate four metallicities [Fe/H], and line types mark three surface
  gravities $\logg$.}
  \label{fig:VvsKurucz}
\end{figure*}

We denote by $I_{\rm Measured}$ our measured intensity profile given by
equation~(\ref{eq:LLD}) with the measured LLDC $u$, normalizing it to unit
r.m.s. intensity. For each theoretical profile from the considered sub-grid we
then compute a scale factor yielding the best agreement with $I_{\rm Measured}$
and mark the rescaled theoretical profile $I_{\rm Kurucz}$. In
\Fig{fig:VvsKurucz} we plot the obtained difference curves
$I_{\rm Kurucz}-I_{\rm Measured}$ as a function of radial position on the stellar
disk for our most precise measurements, i.e., the $\Vs-$ and $\Is$-band
limb darkening. From the plot one can already distinguish by eye the agreement
with Kurucz profiles of different effective temperature, metallicity, and
surface gravity. For instance, all the 4750 K profiles agree with the
measured $\Vs$ profile within 1.3\% of the r.m.s. intensity from disk centre
out to the limb. Nevertheless, in the case of the measured $\Is$ profile it is
hard to judge by eye which group of model profiles agrees better. The 4500~K
models are preferred near the centre, but closer to the limb their accuracy
drops earlier than that of the 4750~K models.

\begin{figure*}[ht]
 \centering
  \includegraphics[width=9cm]{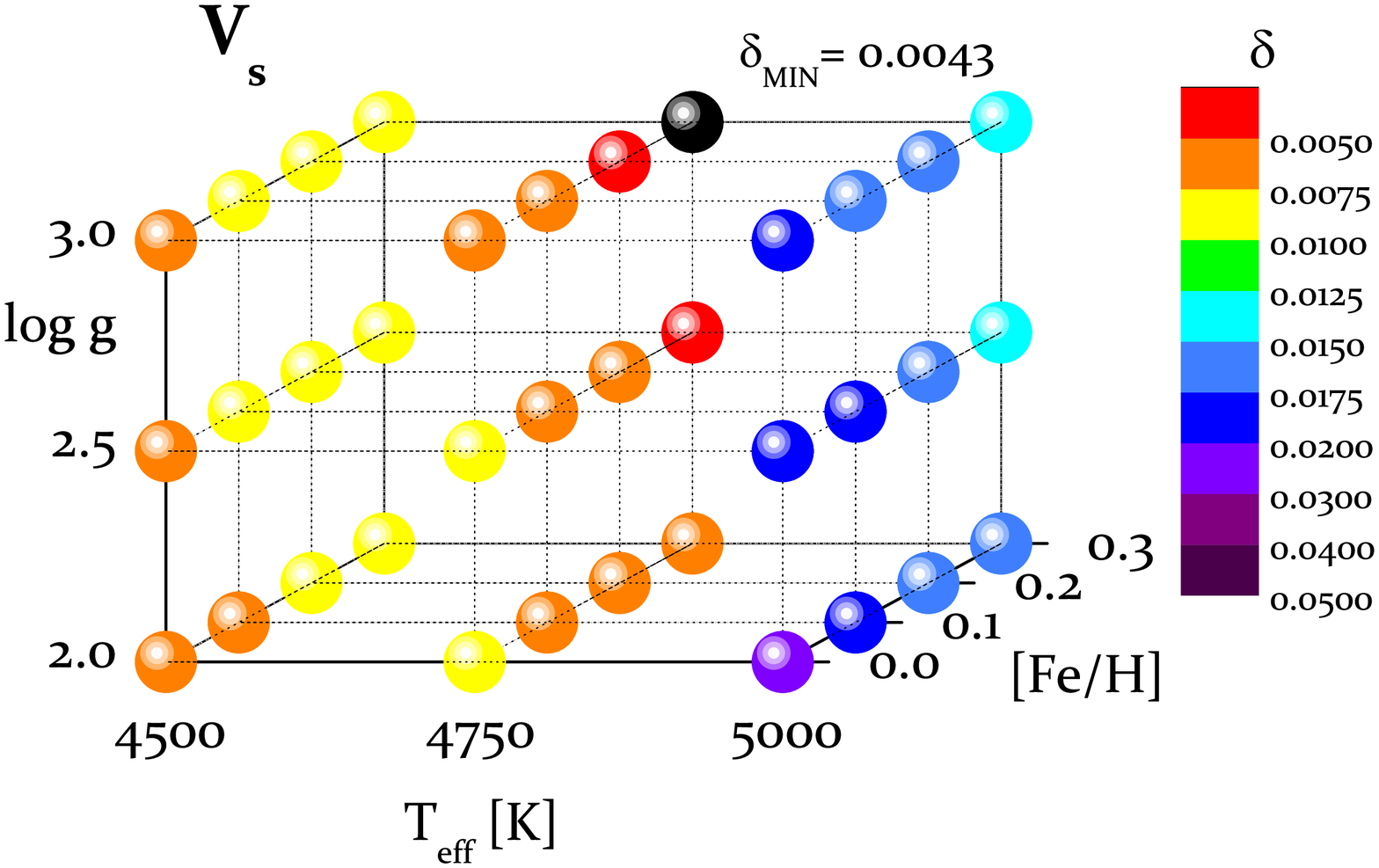}
  \includegraphics[width=9cm]{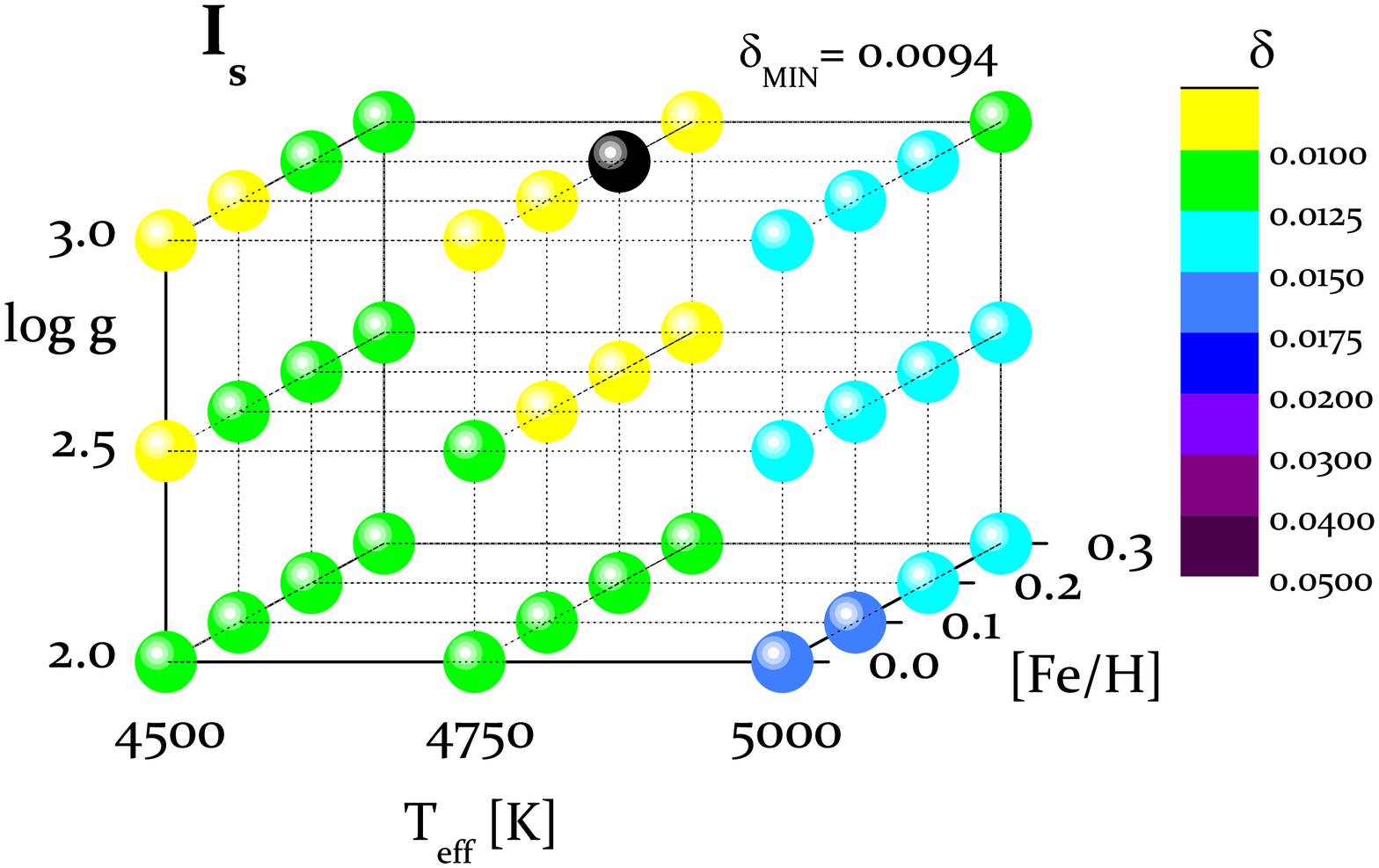}
  \caption{Agreement between the measured $\Vs-$band (left panel) and
  $\Is-$band (right panel) limb-darkening profile and Kurucz limb-darkening
  profiles, as indicated by the colour-coded relative residual $\delta$. With
  the normalization used in \Fig{fig:VvsKurucz}, values of $\delta$ are equal
  to the r.m.s. of each of the difference curves in \Fig{fig:VvsKurucz}. The
  three Cartesian coordinates in the plotted grid correspond to the three main
  stellar-atmosphere parameters in the Kurucz model grid, $\teff$, [Fe/H], and
  $\logg$. The best-agreeing model is marked by the black dot and has the value
  $\delta_{\rm MIN}$ given above the plot.}
  \label{fig:Fitquality}
\end{figure*}

To obtain a quantitative measure of agreement we compute the relative residual
$\delta=\sqrt{\int{(I_{\rm Kurucz}-I_{\rm Measured})^2dr}}/\sqrt{\int{I_{\rm Measured}^2dr}}$,
defined as the r.m.s. difference in units of r.m.s. intensity. The values of
$\delta$ corresponding to the profiles in \Fig{fig:VvsKurucz} are presented
graphically in \Fig{fig:Fitquality}. Each point in the left (right)
grid corresponds to a Kurucz $\Vs$-band ($\Is-$band) profile with the
corresponding $\teff$, [Fe/H], and $\logg$ values as the coordinates.
The colour of the point indicates the value of $\delta$ as shown in the
colour bar. The black point singles out the best agreeing model with the
relative residual $\delta=\delta_{\rm MIN}$ given above the grid. Clearly
$\teff=4750$~K is strongly favoured in both bands, validating our previous
finding. As for the other parameters, there is a weaker preference for higher
[Fe/H] and higher $\logg$.

In order to get an error bar estimate on $\teff$, we perform a similar
analysis for profiles with LLDC values given by the upper and lower error bars
in either band from \Tab{tab:fitparameters}. We find that the lower limits on
$u_{\Vs}$ and $u_{\Is}$ both agree best with 4750~K models, while the upper
limits on both agree best with 4500~K models. Based on these results and the
change of $\delta$ with $\teff$ in \Fig{fig:Fitquality} we conclude that our
limb-darkening measurement yields a temperature estimate of the source star
$\teff=4700^{+100}_{-200}$~K, obtained by comparison with the limb darkening of
Kurucz's models.

The LLDC measured from our analysis for the $I-$band is presented in
\Tab{tab:LDevents} and compared to model LLDC predictions for the given
stellar parameters, together with values for similar giants resolved
by microlensing, namely EROS~2000-BLG-5 from \citet{Fields2003} as reported in
\citet{Yoo2004}, OGLE~2003-BLG-238 from \citet{Jiang2004} and OGLE~2004-BLG-254
from revised fits in \citet{Heyrovsky2008}, which solve discrepancies noted in
\citet{Cassan2006}. Unfortunately, the listed effective temperatures are based
on photometric estimates for all events but OGLE~2004-BLG-254, for which the
spectroscopic measurement (4100~K) disagrees with the photometric one (4500~K)
given in the Table. OGLE~2008-BLG-290 appears to be a twin of EROS~2000-BLG-5,
while OGLE~2003-BLG-238 and OGLE~2004-BLG-254 are slightly hotter.

\begin{table*}[ht]
  \caption{Photometric temperatures and $I-$band limb-darkening coefficients of
  K Bulge giants for OGLE~2008--BLG--290 and other published
  microlensing events with source stars of similar spectral type.}
  \label{tab:LDevents}
  \centering
    \begin{tabular}{lccccccc}
      \hline\hline 
      \multicolumn{1}{c}{Event} & \multicolumn{4}{c}{Source characteristics} &
 Measured LLDC & ATLAS LLDC \\ 
      & Type  & $\mathrm{T_{eff}}$ & $\logg$ & [\element{Fe}/\element{H}] & $u_I$ & $u_I$ \\
      \hline
      & \\
      EROS~2000--BLG--5   & K3~III & 4200~K & $2.3$ & $0.3$ & $0.54^{\mathrm{a}}$ & $0.62$ \\
      OGLE~2003--BLG--238 & K2~III & 4400~K & $2.0$ & $0.0$ & $0.58 \pm 0.06$ & $0.59$ \\
      OGLE~2004--BLG--254 & K3~III & 4500~K & $2.0$ & $0.3$ & $0.62 \pm 0.07$ & $0.59$ \\
      OGLE~2008--BLG--290 & K3~III & 4200~K & $2.5$ & $0.2$ & $0.539 \pm 0.033$ & $0.61$ \\
      & \\
      \hline
    \end{tabular}
\begin{list}{}{}
\item[$^{\mathrm{a}}$] One-parameter fit cannot properly describe the LD profile:
see \Sec{sec:LDmeasure} for details.
\end{list}
\end{table*}

For the present event, the agreement between temperatures estimated from
colours and from limb-darkening coefficients is not satisfactory
($4200 \pm 100$ vs. $4700^{+100}_{-200}$~K, respectively). Unfortunately, we do
not have a spectroscopic estimate of $\teff$, but it is not unusual that
photometric and spectroscopic estimates disagree by several hundred K in cool
giants. We refer to \citet{Affer2005} and \citet{Fulbright2006} for detailed
discussions about this discrepancy.

One way to solve the disagreement is to assume that our target suffers more
extinction than the red clump in the same region. Let us check what it would
imply if the source star suffers 0.1 mag more extinction in the $\Ks-$band than
the clump. According to our adopted reddening law in the near-infrared
(\Equ{eq:RL_JK} and \Equ{eq:RL_HK}), the source star would be at $\Mj=-0.73$
and $\JHo=0.66$, inside the red clump. Adopting typical reddening law ratios of
10 and 5.6 between respectively $\Av$ and $\Ai$ on one side, and $\Aks$ on the
other side, the source star brightens by 0.56 mag in $I$ and becomes bluer by
0.44 mag in $(V-I)$, explaining the observed shifts in the visible CMD with
respect to the red clump centroid not because the star is redder and fainter,
but because it suffers more extinction.

As the typical temperature of Bulge red clump giants is about 4750~K (V. Hill,
personal communication), this would approximately reconcile the effective
temperature deduced from the LD measurements with the photometric temperature.

Obviously, this would slightly decrease (by about 5\%) the angular radius of
the source obtained from the surface brightness colour relations (\Equ{eq:SBVI}
and \Equ{eq:SBVK}) and accordingly all the dependent parameters, but the exact
change depends on the visible to near-infrared extinction ratios, which are not
well known \citep[e.g.][]{Nishiyama2008}.

\section{Lens properties} \label{sec:lens}

With the angular Einstein radius being related to the angular source radius
$\ThS$ as $\ThE = \ThS/\rhoS$, we find $\ThE = 300 \pm 20 \,\muas$. This
enables us to calculate the relative lens-source proper motion:
$\mu = \ThE/\tE = 6.7 \pm 0.4$~mas/yr.  The relative transverse velocity
$\vperp$ between lens and source at the lens distance then follows from
$\vperp = \DL\,\mu$.

From the value of $\ThE$ in mas and $\DS$ in kpc, we obtain a constraint on the
lens mass $\ML$ in solar mass units as
\citep{Dominik1998}:
\begin{equation}
  \ML \,(x) = \frac{\ThE^2 \, \DS}{8.144}\frac{x}{1-x},
\label{eq:mass_x}
\end{equation}
where $x=\DL/\DS$, which for $\DS=7.6$ kpc and $\ThE=0.30$ mas gives:

\begin{equation}
  \ML \,(x) = 0.084 \frac{x}{1-x} \  \msun,
\end{equation}

\begin{figure}[ht]
 \centering
  \includegraphics[width=8.5cm]{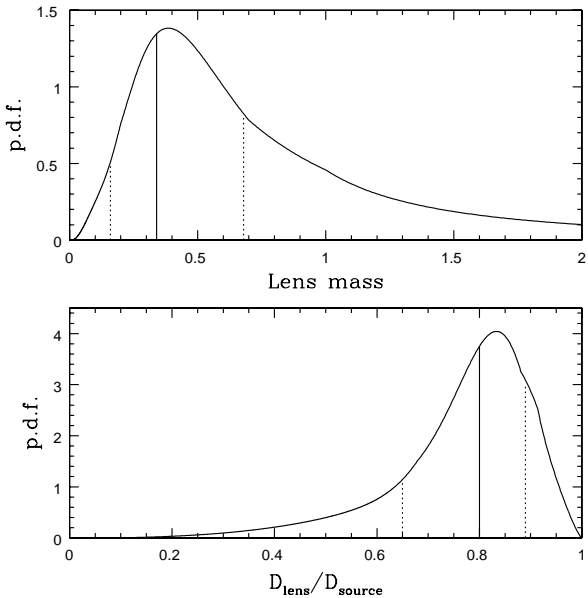}
  \caption{Probability densities for the lens mass $\ML / \msun$ and the lens
   distance $\DL$ from the adopted Galaxy model. Vertical lines correspond to
   median, first and last quartiles.}
  \label{fig:properties}
\end{figure}

In principle, a measurement of the source size in both Einstein radius
and physical units, as well as the measurement of parallax parameters
completely determine the lens location (given the source distance
$\DS$). Indeed, similarly to \Equ{eq:mass_x}, a measured parallax parameter
would provide a relation between lens mass and $x$, but unfortunately the
event is too short ($16$ days $\ll$ 1 year) to provide a
measurement of the orbital parallax, or even to give reasonable limits. For a
similar event (duration $\tE \simeq 13$ days, source size $\rhoS \simeq 0.06$
and very small impact parameter $u_{\circ}$), \cite{Yoo2004} only obtained
a marginal parallax measurement as well.

We then use estimates of the physical parameters, following
\citet{Dominik2006} and assuming his adopted Galaxy model. The event
time-scale $\tE = 16.4~\mbox{days}$ and the angular Einstein radius
$\ThE = 0.3$~mas provide us with probability densities for the lens mass $\ML$
and the lens distance $\DL$, which are shown in \Fig{fig:properties}.

From these, we find a median lens mass $\ML \sim 0.3\,\msun$, a median
projected relative velocity $\vperp \sim 190~\mbox{km}\,\mbox{s}^{-1}$, and
a median lens distance $\DL \sim 6.0~\mbox{kpc}$ for an adopted source
distance of $\DS = 7.6~\mbox{kpc}$. The lens is inferred to reside in the
Galactic Bulge, with $60\,\%$ probability.

\section{Possible planetary deviations} \label{sec:planet}

In this section we include even the 120 points of the SAAO $I-$band
data in the analysis to test whether the overall residuals may harbour the weak
signature of a planetary companion of the lens. The single-lens
extended-source model with fitted limb-darkening coefficients then has
a $\chisq$ of 2485.7 for 2169 data points (after rescaling error bars). We
will now test if a binary lens fit may improve this $\chisq$ by more than some
threshold level. If this were the case, then a planet signal is probably buried
in the small residuals, similarly to the MOA~2007-BLG-400 \citep{Dong2009b} and
MOA~2008-BLG-310 \citep{Janczak2010} events.

Our exploration of the parameter space uses the $q \  w \  \alpha$
parameterization introduced and justified by \citet{Dong2009b}. $q$ is the mass
ratio between the two components of the lens system ($q < 1$), $w$ is the width
of the central caustic as defined by \citet{Chung2005} (given there by their
Eq.~(12), and called vertical width or short diameter) and $\alpha$ is the
trajectory angle with respect to the line joining the two components.

In order to choose the limits of the grid in $\log q$ and $\log w$, we
note that any planetary deviation would be strongly smeared out by the
extended-source effect when the caustic width $w\ll\rhoS$. \citet{Han2009}
argue that a 5\% deviation can be detected even when the central
caustic width $w$ is larger than a quarter of the source
diameter ($0.5 \  \rhoS$). Indeed, with adequate photometric accuracy,
\citet{Dong2009b} clearly detect a planetary deviation in MOA~2007-BLG-400 at
$w = 0.4 \  \rhoS$, and \citet{Batista2009} find some sensitivity even down
to $0.1 \  \rhoS$.

We therefore use a grid with $w$ between $10^{-3}$ and $10^{-2}$ and $q$ between
$10^{-4}$ and $10^{-2}$. We find a marginal improvement of 55 in $\chisq$ in a
region centred around $\log q = -2.5$, $\log w = -2.5$ and $\alpha = -0.459$
rad, corresponding to $w = 0.14 \  \rhoS$. However, the corresponding residuals
do not show any improvement over the single lens case, so we conclude that any
possible planetary deviation is not detectable in this event, given the large
source size.

A similar conclusion can be drawn even for the presence of multiple
planetary companions. Just like in the single-planet case, in a
high-magnification event such as ours lens companions would in effect change
the point-like caustic of the single lens to a small cusped caustic at the same
position. On the light curve this would produce symmetrically placed
perturbations at the times of entering and exiting the limb of the source, a
region where we see no significant systematic residuals.

\section{Other potential systematics} \label{sec:systematics}

In order to affect our limb-darkening measurement and the subsequent
$\teff$ estimate, any potential modeling systematic uncertainty would have to
influence the light-curve shape in the narrow region corresponding to the
caustic crossing. In our case the region is limited to the $\lesssim$
1-day-wide interval around the peak. Having already tested negatively for the
presence of a planetary companion, we turn here to potential uncertainties
related to the source, and to the limb-darkening treatment.

Two relevant source-related uncertainties that could influence the
result are the presence of star spots and the presence of low-level variability
of the source star. The microlensing influence of a star spot would be
temporally limited to the spot-crossing time, and the maximum photometric
effect is given approximately by the fractional radius of the spot
\citep{HS2000}. Judging by the amplitude of the residuals in the peak region,
our event could potentially permit spots with radii $<$ 0.01 stellar radii.
Nevertheless, any such a spot, in addition to being obscured by light-curve
noise, would affect only 0.01 of the source-crossing time. For a longer
duration small-amplitude effect one would be limited to the option of a very
low-contrast larger spot possibly positioned off the projected lens path.
However, such a weak perturbation would be unlikely to affect our results
significantly.

Potential low-level variability of the source should always be
considered, especially because it could go undetected in the absence of the
lens if there is substantial blended light from other stars. In the presence of
the lens it should be noticeable in the residuals near the light-curve peak,
which should also be mutually correlated in the different light curves as seen
in the MACHO Alert 95-30 case \citep{Alcock1997,Heyrovsky2003}. However, only a
source varying significantly on timescales $\lesssim$ the source-crossing time
could potentially affect the limb-darkening measurement. Checking our residuals
in the peak region, we find that they stay below the 1\% level and the
residuals from individual curves do not exhibit a clear correlated pattern.

The results could be potentially affected also by the adopted
limb-darkening treatment. For our measurement we have chosen the linear
limb-darkening law primarily because it is the most widely used
single-parameter law. In addition, it is reasonably accurate for broadband
photometry in the optical part of the spectrum of a wide range of stars. For
the sub-grid of Kurucz models tested in the paper and the relevant photometric
bands, the accuracy of the linear law stays better than its average value for
the full Kurucz grid \citep{Heyrovsky2007}. Within the tested range, the
accuracy is best in the $\Vs-$ band and decreases with effective wavelength to
the $I-$ band. Nevertheless, our $\Vs$-band and $\Is$-band measurements gave us
compatible results. Moreover, in order to compensate for the potential
inadequacy of the law, we compared the measured limb darkening not only with
the LLDCs of the Kurucz models but also directly with their unapproximated
limb-darkening profiles. In principle, one could test more advanced
limb-darkening laws such as the PCA description
\citep{Heyrovsky2003,Heyrovsky2008} or higher-order analytical laws with more
free parameters. However, non-negligible inadequacy of the linear law would be
exhibited by a center-to-limb pattern in the residuals, something we do not see
in our results given the level of light-curve noise. Based on the above we do
not expect the limb-darkening treatment to change our conclusions on the
measured limb darkening and the derived effective temperature of the source
star.

Finally, it should be noted that inaccuracies may arise on the
model-atmosphere side of the comparison. The Kurucz ATLAS9 models used in our
analysis involve various simplifications, such as 1-D radiative transfer in
a plane-parallel atmosphere, the assumption of local thermodynamic equilibrium,
or the mixing-length treatment of convection. Such approximations may affect
the predicted limb-darkening profiles, which reflect the change of physical
conditions in the atmosphere with depth. Accurate measurement of limb darkening
thus provides an opportunity to test the depth structure of model atmospheres
and, hence, the appropriateness of their underlying physical assumptions.

\section{Summary and conclusion} \label{sec:discussion}

We have performed dense photometric monitoring of the microlensing event
OGLE~2008--BLG--290, a short duration and very small impact parameter
microlensing event generated by a point-like lens transiting a giant star. The
peak magnification was about $A(\to)\sim 100$, effectively multiplying
the diameter of our network telescopes by a factor $\sim 10$. Using a calibrated
colour-magnitude diagram analysis and isochrones, we find a source
angular radius $\ThS \simeq 6.5 \pm 0.4 \,\muas$, and a physical radius
$\RS \simeq 10 \ \rsun$ at the adopted source distance $\DS = 7.6$~kpc. A
Galaxy model together with the event time-scale $\tE = 16.4~\mbox{days}$ and
the angular Einstein radius $\ThE = 0.3$~mas yielded statistically inferred
median parameters, namely a lens mass $M\sim 0.3\,\msun$, a lens distance
$\DL \sim 6.0~\mbox{kpc}$, and a relative velocity
$\vperp \sim 190~\mbox{km}\,\mbox{s}^{-1}$.

From our photometric data, we have derived accurate measurements of the
linear limb-darkening coefficients of the source for the $\Vs$, SDSS
$r$, $R$, MOA-red, $\Is$, and $I$ broadband filters. The obtained
limb-darkening profiles lead to a source-star temperature estimate of
$\teff=4700^{+100}_{-200}$~K when compared with the limb-darkening profiles of
Kurucz's ATLAS9 atmosphere models. This result is in marginal disagreement with
the corresponding estimate based on measured colours, a finding that has
already been noted for several previous events measured by the microlensing
technique, which also parallels the known discrepancies in temperature
estimates based on photometry vs. spectroscopy of similar cool giants
\citep{Fulbright2006}. Possible explanations involve reddening corrections
which may lead to wrong colour estimates, or inaccuracies in model atmosphere
physics.

\begin{acknowledgements}

We express our gratitude to ESO for a two months invitation at Santiago
headquarters, Chile, where part of this work was achieved.
We are very grateful to the observatories that support our science
(Bronberg, Canopus, CTIO, ESO, IRSF, LCOGT, Liverpool, LOAO, MOA, OGLE, Perth, 
SAAO, Skinakas) via the generous allocation of time that makes this work 
possible.  The operation of Canopus Observatory is in part supported by a 
financial contribution from David Warren. The OGLE project is partially 
supported by the Polish MNiSW grant N20303032/4275 to AU.
Allocation of the Holmes grant from the French Agence Nationale de la
Recherche has been indispensable to finance observing trips and meeting travels,
and is gratefully acknowledged here. DH was supported by Czech Science
Foundation grant GACR 205/07/0824 and by the Czech Ministry of Education
project MSM0021620860. CH was supported by the grant 2009-0081561 of National
Research Foundation of Korea. TCH was financed for his astronomical research at
the Armagh Observatory by the Department for Culture, Arts and Leisure,
Northern Ireland, UK. DR and JS acknowledge support from the Communaut\'e
fran\c{c}aise de Belgique - Actions de recherche concert\'ees - Acad\'emie
universitaire Wallonie-Europe. PF wishes to thank Noriyuki Matsunaga for
discussions about the interplay between adopted distance and derived
extinction, David Nataf for measuring the red giant clump in the
recently released OGLE-III photometric catalogue, and Etienne Bachelet for
checking the whole chain from extinction to source size using a refined
method. We are grateful to the anonymous referee for constructive comments that
helped us improve the manuscript.

This publication makes use of data products from the 2MASS project,
as well as the SIMBAD database, Aladin and Vizier catalogue
operation tools (CDS Strasbourg, France). The Two Micron All Sky
Survey is a joint project of the University of Massachusetts and the
Infrared Processing and Analysis Center/California Institute of
Technology, funded by the National Aeronautics and Space
Administration and the National Science Foundation.

\end{acknowledgements}

\bibliographystyle{aa}
\bibliography{biblio_290}
\end{document}